\pdfoutput=1 
\documentclass[conference]{IEEEtran}
\IEEEoverridecommandlockouts
\usepackage{cite}
\usepackage{amsmath,amssymb,amsfonts}
\usepackage{algorithmic}
\usepackage{graphicx}
\usepackage[caption=false,font=footnotesize]{subfig} 
\usepackage{textcomp}
\usepackage{xcolor}
\usepackage{etoolbox}
\usepackage{tikz}
\usepackage{booktabs}
\usetikzlibrary{svg.path}
 
\graphicspath{{figures/}{figures_jul10/}}
\usepackage[capitalize,noabbrev]{cleveref}
    \crefname{figure}{Fig.}{Figs.}
    \crefname{subfigure}{Fig.}{Figs.}
    \crefname{equation}{}{}
    \crefname{section}{Section}{Sections}
    \crefname{table}{Table}{Tables}
\DeclareMathOperator{\sgn}{sgn}
\usepackage[group-minimum-digits=4,per-mode=fraction]{siunitx}
\makeatletter
\newcommand{\IEEEleftcaption}[2]{%
  \begingroup
  \long\def\@makecaption##1##2{%
    \@IEEEfigurecaptionsepspace
    \setbox\@tempboxa\hbox{\normalfont\footnotesize {##1.}\nobreakspace\nobreakspace}%
    \parbox[t]{\hsize}{\normalfont\footnotesize\noindent\unhbox\@tempboxa ##2}%
  }%
  \caption{#1}%
  \label{#2}%
  \endgroup
}
\makeatother
 
\makeatletter
\patchcmd{\@maketitle}
  {\addvspace{0.5\baselineskip}\egroup}
  {\addvspace{-0.3\baselineskip}\egroup}
  {}
  {}
\makeatother
\definecolor{orcidlogocolor}{HTML}{A6CE39}
\def\orcidsymbolpathouter{svg{M256,128c0,70.7-57.3,128-128,128C57.3,256,0,198.7,0,128C0,57.3,57.3,0,128,0C198.7,0,256,57.3,256,128z}}
\def\orcidsymbolpathinner{svg{M86.3,186.2H70.9V79.1h15.4v48.4V186.2z}
                     svg{M108.9,79.1h41.6c39.6,0,57,28.3,57,53.6c0,27.5-21.5,53.6-56.8,53.6h-41.8V79.1z M124.3,172.4h24.5c34.9,0,42.9-26.5,42.9-39.7c0-21.5-13.7-39.7-43.7-39.7h-23.7V172.4z}
                     svg{M88.7,56.8c0,5.5-4.5,10.1-10.1,10.1c-5.6,0-10.1-4.6-10.1-10.1c0-5.6,4.5-10.1,10.1-10.1C84.2,46.7,88.7,51.3,88.7,56.8z}}
\tikzset{
    orcidlogo/.pic={
        \fill[orcidlogocolor] \orcidsymbolpathouter;
        \fill[white] \orcidsymbolpathinner;
    },
}

\def\orcid#1{\tikz[yscale = -1, scale = 0.035, transform shape, baseline = -7.5pt] \pic{orcidlogo};\,orcid.org/#1}
 
\def\BibTeX{{\rm B\kern-.05em{\sc i\kern-.025em b}\kern-.08em
    T\kern-.1667em\lower.7ex\hbox{E}\kern-.125emX}}
\newcommand{\paperdoi}{}
\usepackage{eso-pic}
\newcommand{\copyrightnotice}{%
    \AddToShipoutPictureBG*{%
        \AtPageLowerLeft{%
            \put(\LenToUnit{0.5\paperwidth},\LenToUnit{0.32in}){%
                \makebox[0pt][c]{\parbox[b]{\dimexpr\paperwidth-1.25in\relax}{\centering\scriptsize
                    Accepted version of a paper presented at the 2026 IEEE Energy Conversion Congress and Expo (ECCE), Vancouver, Canada.%
                    \if\relax\detokenize\expandafter{\paperdoi}\relax\else\ DOI: \paperdoi.\fi\\
                    \copyright\,2026 IEEE. Personal use of this material is permitted. Permission from IEEE must be obtained for all other uses, in any current or future media, including reprinting/republishing this material for advertising or promotional purposes, creating new collective works, for resale or redistribution to servers or lists, or reuse of any copyrighted component of this work in other works.}}}}}}
 
\begin{document}
\copyrightnotice
 
\title{A Modulation Strategy for Bidirectional Integrated Onboard and Wireless EV Charging
    \thanks{This work was supported by the Natural Sciences and Engineering Research Council of Canada (NSERC) under Discovery Grant RGPIN-2024-04642, by an NSERC Canada Graduate Research Scholarship - Master's (CGRS-M), and by the University of Manitoba Graduate Fellowship (UMGF).}
}
 
\author{\IEEEauthorblockN{Manveer Aujla}
    \IEEEauthorblockA{\textit{Electrical and Computer Engineering} \\
        \textit{University of Manitoba}\\
        Winnipeg, Canada \\
        aujlam@myumanitoba.ca\\
        \orcid{0009-0006-9898-140X}}
    \and
    \IEEEauthorblockN{Gavin Feher}
    \IEEEauthorblockA{\textit{Electrical and Computer Engineering} \\
        \textit{University of Manitoba}\\
        Winnipeg, Canada \\
        feherg@myumanitoba.ca\\
        \orcid{0009-0008-1021-2898}}
    \and
    \IEEEauthorblockN{Caniggia Viana}
    \IEEEauthorblockA{\textit{Electrical and Computer Engineering} \\
        \textit{University of Manitoba}\\
        Winnipeg, Canada \\
        caniggia.viana@umanitoba.ca\\
        \orcid{0000-0001-8550-9428}}
}
 
\maketitle

\begin{abstract}
    Integrated electric vehicle chargers reduce cost by sharing power conversion hardware between the onboard charger (OBC) and the wireless charger (WC). A previously proposed dual active bridge (DAB)-based OBC/WC shares the rectification stage between the OBC and WC; however, the modulation required to realize its wireless operation has not yet been reported. This paper details a modulation strategy that enables the operation of the integrated OBC/WC topology. Differential-mode operation is used for the OBC, and common-mode operation is used for the WC. In OBC mode, the shared full-bridge produces differential-mode voltage for wired power transfer while ideally generating zero common-mode voltage on the wireless charging path. In WC mode, the same full-bridge produces common-mode voltage for wireless power transfer while ideally generating zero differential-mode voltage on the onboard charging path. The method preserves conventional DAB phase-shift power control and enables bidirectional power transfer in wired and wireless charging modes without additional contactors. Design considerations for the OBC transformer and WC coils are derived based on the resulting square-wave voltage amplitudes. Offline PLECS simulations demonstrate power tracking and decoupled charger operation at ±\SI{7.4}{\kilo\watt}. Hardware-in-the-loop simulations using a PLECS RT Box and an external Texas Instruments microcontroller further validate the proposed modulation method.
\end{abstract}


\section{Introduction}
High-frequency galvanically-isolated dc/dc converters are widely used in electric vehicle (EV) onboard chargers (OBCs) for their efficiency and power density \cite{2017trans}. Wireless chargers (WCs) provide a convenient alternative by allowing charging without physical connectors. Both OBCs and WCs require dedicated in-vehicle power stages to interface with the traction battery, thereby increasing system cost. To address this, numerous integrated topologies have been proposed that combine hardware from the OBC and WC. Some topologies use a diode bridge for rectification in the WC \cite{2024PEtrans}, but this precludes bidirectional wireless charging. Others feature an active full-bridge in the WC, suggesting bidirectional wireless charging capability; however, the reported operation is limited to grid-to-vehicle charging \cite{2021IAtrans, 2024IEtrans, 2025PEtrans, 2025PEtrans2}. Furthermore, some topologies require additional mechanical switches to change between operating modes, adding cost and complexity \cite{2024IEtrans, 2025PEtrans}.

\begin{figure*}[htbp]
    \centering
    \includegraphics[width=1\textwidth]{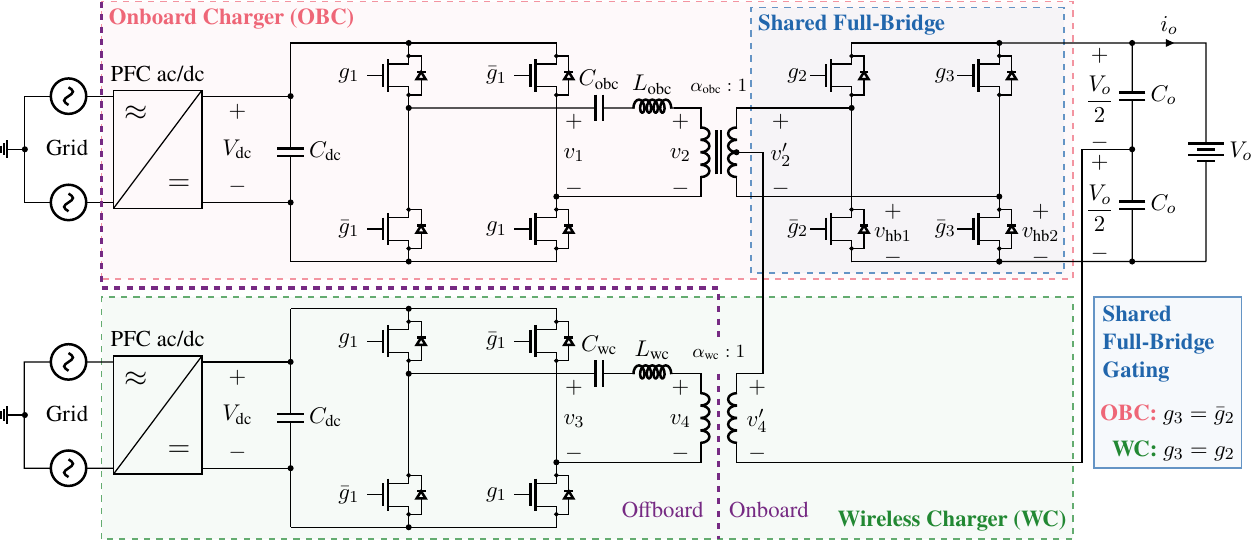}
    \caption{Topology of the DAB-based integrated OBC/WC system.}
    \label{fig:topology}
\end{figure*}

The authors of \cite{patent} have proposed several integrated topologies that are capable of bidirectional wired and wireless charging without additional contactors. These topologies leverage differential-mode and common-mode operation to ensure that only one charging system is energized at a time. One proposed design in \cite{patent} is the dual active bridge (DAB)-based integrated OBC/WC shown in \cref{fig:topology}. The OBC transfers power using a differential-mode voltage, an approach studied extensively in the literature. In contrast, the WC requires common-mode voltage to transfer power. The modulation required to realize this operation was not described in \cite{patent}, and a modulation strategy enabling this topology has not yet been reported.

This paper introduces a novel modulation scheme allowing independent control of the OBC and WC using differential-mode and common-mode voltage, respectively. For the first time, bidirectional power flow for wired and wireless charging can be demonstrated for the integrated OBC/WC of \cite{patent}.

\section{Non-Integrated OBC/WC Overview}
Conventional non-integrated EV chargers use separate wired and wireless charging paths that interface with the battery \cite{2012TVT}. For some bidirectional architectures, these separate paths are implemented as two DAB converters. Each DAB consists of two full-bridge converters connected by a high-frequency transformer (or wireless coils in the case of a WC), usually with capacitive compensation. When using single-phase shift (SPS) modulation, the diagonal switches in each full-bridge are driven together with a 50\% duty cycle. The relative phase shift $\phi$ of the resulting square-wave voltages dictates the power transfer magnitude and direction \cite{1991trans}. Maximum grid-to-vehicle power transfer occurs when $\phi = \frac{\pi}{2}$ (battery-side full-bridge lags), and maximum vehicle-to-grid power transfer occurs when $\phi = -\frac{\pi}{2}$ (battery-side full-bridge leads). Hence, the DAB can be operated bidirectionally by controlling the phase shift in the range $ -\frac{\pi}{2} \leq \phi \leq \frac{\pi}{2} $. Although this topology is modular, it contains redundant hardware, such as the two battery-side full-bridges. Removing the redundant hardware reduces cost, motivating an integrated OBC/WC.

\section{Integrated OBC/WC Topology and Proposed Modulation Strategy}
Consider the DAB-based integrated OBC/WC in \cref{fig:topology} with a shared full-bridge at the battery side. For simplicity, the leakage inductances ($L_\text{obc}$ and $L_\text{wc}$) and compensation capacitances ($C_\text{obc}$ and $C_\text{wc}$) have been referred to the primary side of the OBC transformer and WC coils. This design decouples OBC and WC operation without additional contactors by leveraging an added electrical path in a centre-tapped transformer and split output capacitors. To verify this decoupling, it must be demonstrated that the OBC can transfer power with negligible voltage on the WC side, and the WC can transfer power with negligible voltage on the OBC side.

\subsection{OBC Mode}
\begin{figure}[htbp]
    \centering
    \includegraphics[width=0.8\linewidth]{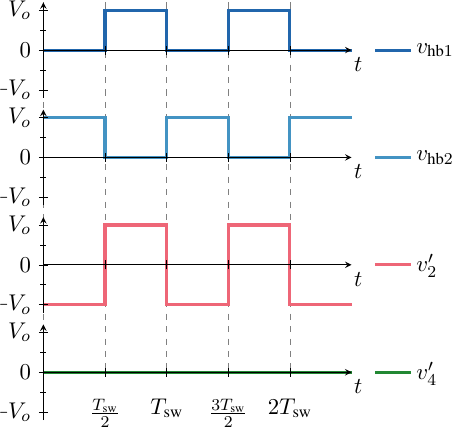}
    \caption{Waveforms for operating the system of \cref{fig:topology} in OBC modulation mode where $g_3=\bar{g}_2$.}
    \label{fig:OBC_analysis_plots}
\end{figure}

The OBC full-bridge can produce a square-wave voltage $v_1$, so if the shared full-bridge can produce a square-wave voltage $v_2'$, power can be transferred by controlling their relative phase shift. To avoid energizing the WC coils, the shared full-bridge should produce negligible voltage on the WC coils, meaning $v_4'=0$. $v_2'$ and $v_4'$ can be expressed as a differential-mode and common-mode voltage, respectively. By KVL,
\begin{equation}\label{eq:DM}
    v_2' = v_{\text{hb}1} - v_{\text{hb}2}.
\end{equation}
Assuming equal impedance on both sides of the centre-tap,
\begin{equation}\label{eq:CM}
    v_4'= \frac{1}{2}v_{\text{hb}1} + \frac{1}{2}v_{\text{hb}2} - \frac{V_o}{2},
\end{equation}
where $v_{\text{hb}1}$ and $v_{\text{hb}2}$ are square waves from $0$ to $V_o$. In terms of the signum function,
\begin{equation}\label{eq:vhb1}
    v_{\text{hb}1}
    = \frac{V_o}{2} + \frac{V_o}{2}\sgn\bigl[\sin(\omega t)\bigr],
\end{equation}
and
\begin{equation}\label{eq:vhb2obc}
    v_{\text{hb}2}
    = \frac{V_o}{2} - \frac{V_o}{2}\sgn\bigl[\sin(\omega t)\bigr].
\end{equation}
In conventional SPS modulation, $g_3 = \bar{g}_2$; hence, $v_{\text{hb}1}$ and $v_{\text{hb}2}$ are \SI{180}{\degree} apart as reflected in \eqref{eq:vhb1} and \eqref{eq:vhb2obc}. By substituting \eqref{eq:vhb1} and \eqref{eq:vhb2obc} into \eqref{eq:DM} and \eqref{eq:CM},
\begin{equation}\label{eq:v2obc}
    v_2'
    = V_o \sgn\bigl[\sin(\omega t)\bigr],
\end{equation}
and
\begin{equation}\label{eq:v4obc}
    \begin{aligned}
        v_4'
         & = \frac{1}{2}\biggl(
        \frac{V_o}{2} + \frac{V_o}{2}\sgn\bigl[\sin(\omega t)\bigr]\biggr)                 \\
         & + \frac{1}{2}\biggl(\frac{V_o}{2} - \frac{V_o}{2}\sgn\bigl[\sin(\omega t)\bigr]
        \biggr)
        - \frac{V_o}{2}                                                                    \\
         & = 0.
    \end{aligned}
\end{equation}
Thus, as illustrated in \cref{fig:OBC_analysis_plots}, $v_2'$ is a square wave with amplitude $V_o$ and $v_4'=0$.

\subsection{WC Mode}
The WC full-bridge can produce a square-wave voltage $v_3$, so if the shared full-bridge can produce a square-wave voltage $v_4'$, power transfer can again be controlled by their relative phase shift. Unfortunately, conventional SPS modulation does not work for WC mode. Regardless of the states of $v_1$ and $v_3$, when the shared full-bridge is modulated with $g_3 = \bar{g}_2$, it produces a square wave $v_2'$ and $v_4'=0$, as defined by \eqref{eq:v2obc} and \eqref{eq:v4obc}, respectively. This means that power cannot be transferred via the WC, as the common-mode voltage would be zero. Furthermore, there would be a large voltage on the OBC side, which may undesirably energize the OBC. A new modulation scheme is required to enable WC functionality.

\begin{figure}[htbp]
    \centering
    \includegraphics[width=0.8\linewidth]{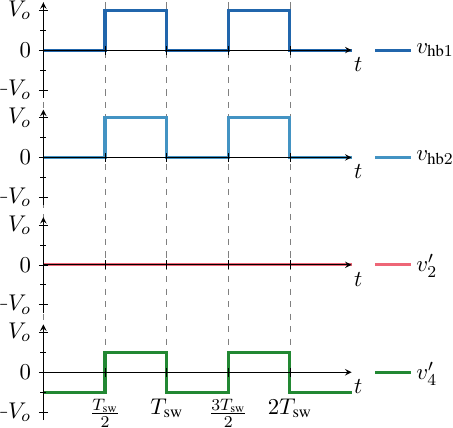}
    \caption{Waveforms for operating the system of \cref{fig:topology} in WC modulation mode where $g_3=g_2$.}
    \label{fig:WC_analysis_plots}
    \vspace{-1.0em}
\end{figure}

The proposed modulation strategy is to drive the top two switches of the shared full-bridge together; that is, $g_3=g_2$. This implies
\begin{equation}\label{eq:vhb12wpt}
    v_{\text{hb}1} = v_{\text{hb}2}
    = \frac{V_o}{2} + \frac{V_o}{2}\sgn\bigl[\sin(\omega t)\bigr].
\end{equation}
Voltages $v_2'$ and $v_4'$ can be found by substituting \eqref{eq:vhb12wpt} into \eqref{eq:DM} and \eqref{eq:CM}, respectively:
\begin{equation}\label{eq:v2wpt}
    v_2' = 0,
\end{equation}
\begin{equation}\label{eq:v4wpt}
    \begin{aligned}
        v_4'
         & = \frac{1}{2}\biggl(
        \frac{V_o}{2} + \frac{V_o}{2}\sgn\bigl[\sin(\omega t)\bigr]\biggr)                 \\
         & + \frac{1}{2}\biggl(\frac{V_o}{2} + \frac{V_o}{2}\sgn\bigl[\sin(\omega t)\bigr]
        \biggr)
        - \frac{V_o}{2}                                                                    \\
         & = \frac{V_o}{2}\sgn\bigl[\sin(\omega t)\bigr].
    \end{aligned}
\end{equation}
Hence, as illustrated in \cref{fig:WC_analysis_plots}, $v_4'$ is a square wave with amplitude $\frac{V_o}{2}$ and $v_2'=0$. Therefore, the proposed modulation strategy, where $g_3=g_2$, achieves what conventional SPS modulation could not: it produces a common-mode voltage that can be used to operate the WC while simultaneously de-energizing the OBC.

\subsection{Turns Ratio and Output Capacitance Selection}
To minimize reactive power flow in OBC mode, the amplitudes of $v_1$ and $v_2$ should be equal at the nominal design point. From \eqref{eq:v2obc}, $v_2'$ has an amplitude of $V_o$. Referring this voltage to the primary side of the OBC transformer, $v_2$ has an amplitude of $\alpha_\text{obc}V_o$. The amplitude of $v_1$ is $V_\text{dc}$, and the OBC transformer turns ratio can be derived by equating the two amplitudes and rearranging for $\alpha_\text{obc}$:
\begin{equation}
    \alpha_\text{obc} = \frac{V_\text{dc}}{V_o}.
\end{equation}

Likewise, in WC mode, the amplitudes of $v_3$ and $v_4$ should be equal at the nominal design point. From \eqref{eq:v4wpt}, $v_4'$ has an amplitude of $\frac{V_o}{2}$. Assuming a transformer-based model for the WC coils with an equivalent turns ratio of $\alpha_\text{wc}$, $v_4'$ can be referred to the primary side to show that $v_4$ has an amplitude of $\frac{\alpha_\text{wc}V_o}{2}$. The amplitude of $v_3$ is $V_\text{dc}$, and the equivalent turns ratio of the WC coils can be derived by equating the two amplitudes and rearranging for $\alpha_\text{wc}$:
\begin{equation}
    \alpha_\text{wc} = \frac{2V_\text{dc}}{V_o}.
\end{equation}
The WC coil equivalent turns ratio is therefore double that of the OBC transformer turns ratio; this is because $v_4'$ is half the amplitude of $v_2'$. Since $V_o$ is the battery voltage, it varies with state of charge. Thus, $\alpha_\text{obc}$ and $\alpha_\text{wc}$ should be designed for the most common operating battery voltage to maximize efficiency.

In WC mode, the output capacitors $C_o$ are part of the electrical path containing $L_\text{wc}$ and $C_\text{wc}$. Hence, to keep the equivalent impedance inductive and to stiffen $V_o$,
\begin{equation}
    C_o \gg \frac{1}{(2\pi f_{\text{sw,wc}})^2 L_\text{wc}}.
\end{equation}
Another capacitor may be added in parallel with $V_o$ to further stiffen the voltage if desired.

\begin{figure*}[htbp]
    \centering
    \includegraphics[width=0.85\textwidth]{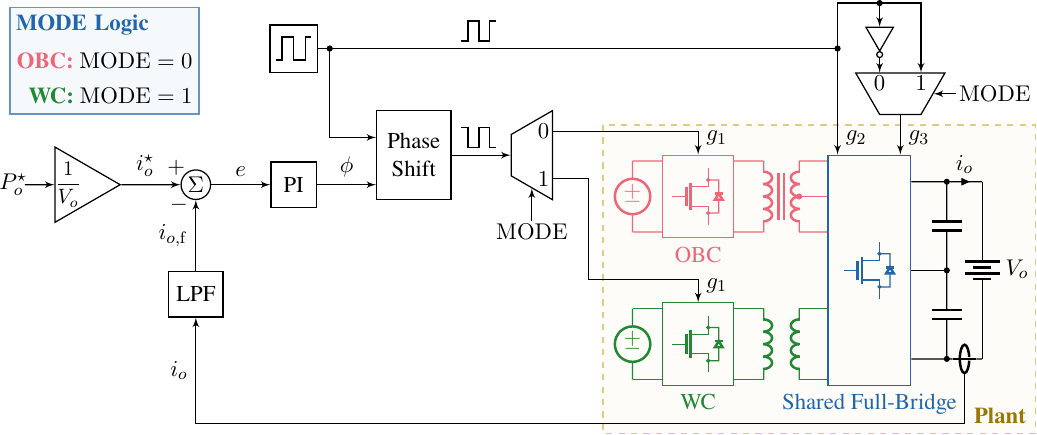}
    \caption{Control diagram of the DAB-based integrated OBC/WC system.}
    \label{fig:controlDiagram}
    \vspace{-1.0em}
\end{figure*}

\section{Control Method}

A control block diagram with the proposed modulation technique is shown in \cref{fig:controlDiagram}. The battery current $i_o$ is sensed, filtered, and compared with the reference current $i_o^\star$. The reference current is obtained by dividing the commanded power transfer $P_o^\star$ by $V_o$, which keeps the current reference consistent if the battery voltage changes. The resulting current error $e$ is fed into a proportional-integral (PI) controller to generate the phase shift $\phi$. Two square-wave gating signals are then produced: $g_2$ is fed to the shared full-bridge, while $g_1$ is phase-shifted by $\phi$ and applied to either the OBC or WC full-bridge according to the selector bit $\mathrm{MODE}$. Diagonal switches in the OBC and WC full-bridges are always driven together. In contrast, the shared full-bridge is driven according to the proposed modulation method: $g_3=\bar{g}_2$ for OBC mode ($\mathrm{MODE} = 0$) and $g_3=g_2$ for WC mode ($\mathrm{MODE} = 1$). This arrangement preserves conventional DAB phase-shift control while selecting which subsystem is energized.

\section{Offline Simulation}

Simulations were performed in PLECS to validate the proposed modulation scheme and control structure. A reference of $P_o^\star=\SI{7.4}{\kilo\watt}$ was used for the OBC, a typical amount of power for level-2 charging \cite{2021access}. For the WC, $P_o^\star=\SI{-7.4}{\kilo\watt}$ was used to demonstrate wireless vehicle-to-grid power transfer. These two cases are representative of the system's performance; for brevity, wired vehicle-to-grid power transfer and wireless grid-to-vehicle power transfer are not shown in this paper. \cref{tab:simParams} lists the offline simulation parameters, and the results are shown in \cref{fig:Plots_OBC_Offline} and \cref{fig:Plots_WC_Offline}.

\begin{table}[htbp]
    \centering
    \caption{Offline Simulation Parameters.}
    \label{tab:simParams}
    \renewcommand{\arraystretch}{1.05}
    \small
    \begin{tabular}{
        >{\raggedright\arraybackslash}p{4.0cm} c @{\hspace{0.5cm}} c
        }
        \toprule
        \textbf{Parameter}           & \textbf{Symbol}     & \textbf{Value}                             \\
        \midrule
        \multicolumn{3}{l}{\textbf{System}}                                                             \\
        \addlinespace
        DC Voltage                   & $V_\text{dc}=V_o$   & \SI{400}{\volt}                            \\
        Output Capacitance           & $C_o$               & \SI{22}{\micro\farad}                      \\
        Controller Proportional Gain & $K_p$               & \num{0.5}\si{\degree\per\ampere}           \\
        Controller Integral Gain     & $K_i$               & \num{50}\si{\degree\per\ampere\per\second} \\
        \midrule
        \multicolumn{3}{l}{\textbf{Onboard Charger (OBC)}}                                              \\
        \addlinespace
        Switching Frequency          & $f_\text{sw,obc}$   & \SI{100}{\kilo\hertz}                      \\
        Turns Ratio                  & $\alpha_\text{obc}$ & 1                                          \\
        Leakage Inductance           & $L_\text{obc}$      & \SI{34}{\micro\henry}                      \\
        Compensation Capacitance     & $C_\text{obc}$      & \SI{135}{\nano\farad}                      \\
        \midrule
        \multicolumn{3}{l}{\textbf{Wireless Charger (WC)}}                                              \\
        \addlinespace
        Switching Frequency          & $f_\text{sw,wc}$    & \SI{85}{\kilo\hertz}                       \\
        Turns Ratio                  & $\alpha_\text{wc}$  & 2                                          \\
        Leakage Inductance           & $L_\text{wc}$       & \SI{218.4}{\micro\henry}                   \\
        Compensation Capacitance     & $C_\text{wc}$       & \SI{17.5}{\nano\farad}                     \\
        \bottomrule
    \end{tabular}
\end{table}

\begin{figure}[htbp]
    \centering
    \subfloat[Transient responses: (i) power tracking and (ii) phase shift.\label{fig:Offline_OBC_transient}]{%
        \begin{minipage}{1\linewidth}
            \centering
            \includegraphics[width=1\linewidth]{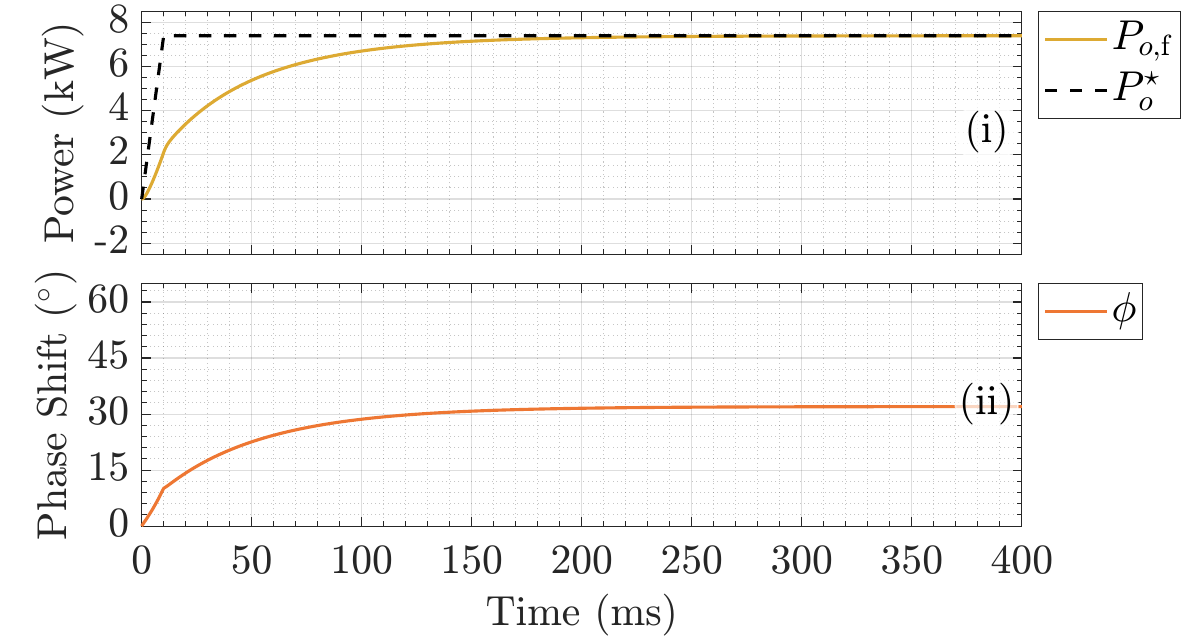}
        \end{minipage}%
    }
    \par
    \subfloat[Decoupled operation: (i) OBC voltages, (ii) WC voltages, (iii) OBC full-bridge gating, and (iv) shared full-bridge gating.\label{fig:Offline_OBC_quick}]{%
        \begin{minipage}{1\linewidth}
            \centering
            \includegraphics[width=1\linewidth]{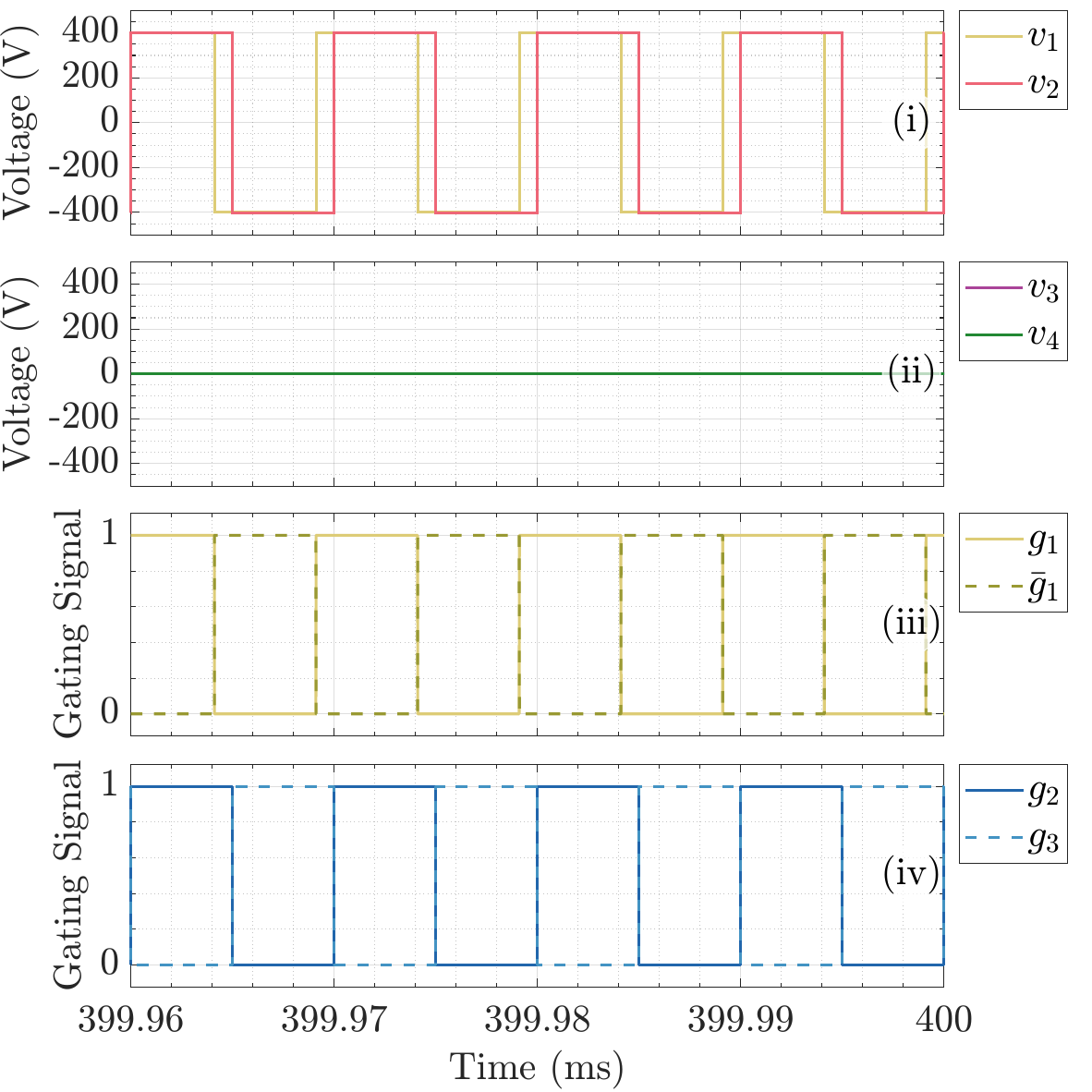}
        \end{minipage}%
    }
    \par
    \IEEEleftcaption{Offline simulation results tracking $P_o^\star=\SI{7.4}{\kilo\watt}$ in OBC mode.}{fig:Plots_OBC_Offline}
    \vspace{-1.5em}
\end{figure}

\begin{figure}[htbp]
    \centering
    \subfloat[Transient responses: (i) power tracking and (ii) phase shift.\label{fig:Offline_WC_transient}]{%
        \begin{minipage}{1\linewidth}
            \centering
            \includegraphics[width=1\linewidth]{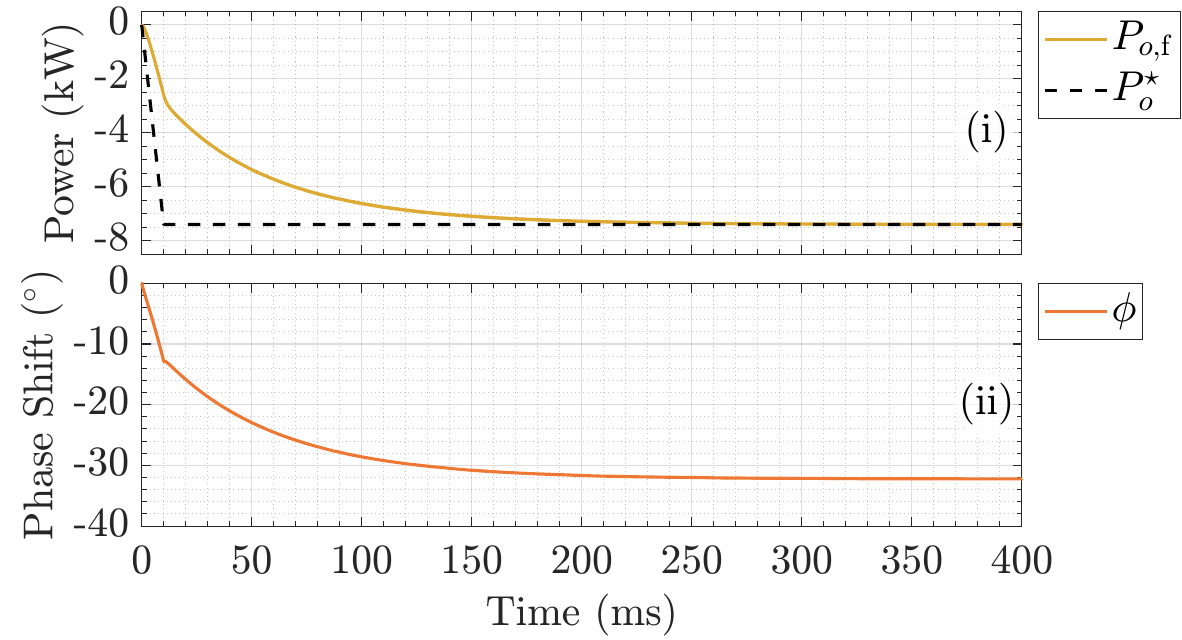}
        \end{minipage}%
    }
    \par
    \subfloat[Decoupled operation: (i) OBC voltages, (ii) WC voltages, (iii) WC full-bridge gating, and (iv) shared full-bridge gating.\label{fig:Offline_WC_quick}]{%
        \begin{minipage}{1\linewidth}
            \centering
            \includegraphics[width=1\linewidth]{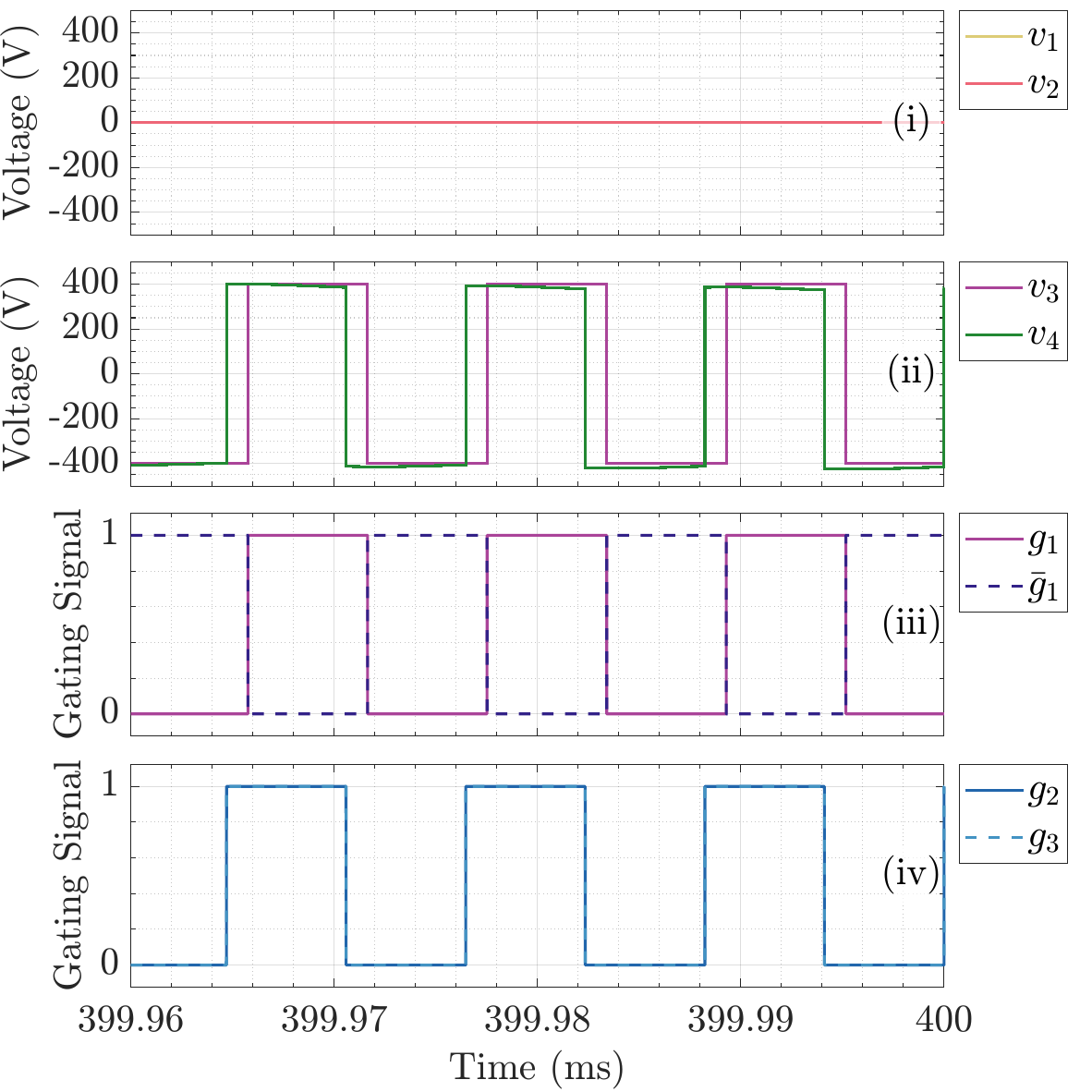}
        \end{minipage}%
    }
    \par
    \IEEEleftcaption{Offline simulation results tracking $P_o^\star=\SI{-7.4}{\kilo\watt}$ in WC mode.}{fig:Plots_WC_Offline}
    \vspace{-1.6em}
\end{figure}

The offline simulation results in \cref{fig:Offline_OBC_transient} and \cref{fig:Offline_WC_transient} show that the controller tracks the reference for OBC grid-to-vehicle and WC vehicle-to-grid operation, respectively. Hence, bidirectional power transfer has been demonstrated using the proposed modulation strategy. Furthermore, the voltages are very similar to those derived analytically: \cref{fig:Offline_OBC_quick} (i)--(ii) show that $v_2$ is a square wave and $v_4$ is nearly zero in OBC mode, while \cref{fig:Offline_WC_quick} (i)--(ii) show that $v_4$ is a square wave and $v_2$ is nearly zero in WC mode. A microvolt-range voltage appears on the inactive charger in each mode; this voltage is unlikely to significantly energize the inactive charger or cause noticeable losses. The square-wave amplitudes are equal in both modes when referred to the same side, validating the selected turns ratios for the OBC transformer and WC coils at the nominal battery voltage used in these simulations.

In OBC mode, the gating signals in the shared full-bridge are \SI{180}{\degree} apart, as shown in \cref{fig:Offline_OBC_quick} (iv). Conversely, in WC mode, the gating signals are in phase, as shown in \cref{fig:Offline_WC_quick} (iv). Taken together with the generated voltages shown in \cref{fig:Offline_OBC_quick} (i)--(ii) and \cref{fig:Offline_WC_quick} (i)--(ii), the analytical derivation shown in \cref{fig:OBC_analysis_plots} and \cref{fig:WC_analysis_plots} has been confirmed in simulation. Finally, \cref{fig:Offline_OBC_quick} (iii) and \cref{fig:Offline_WC_quick} (iii) show that the OBC and WC full-bridges are driven conventionally in both modes.

\section{Hardware-in-the-Loop Simulation}
\begin{figure}[htbp]
    \vspace{-0.7em}
    \centering
    \includegraphics[width=1\linewidth]{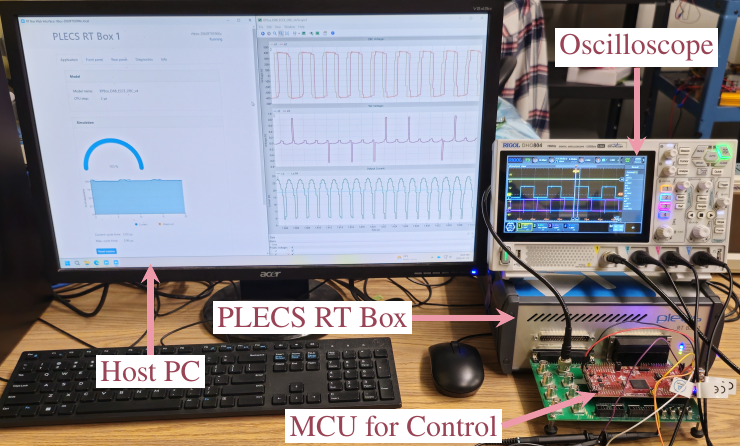}
    \vspace{-1.8em}
    \caption{HIL simulation setup. The RT Box models the plant, the MCU executes the control and provides gating signals, and the oscilloscope and host PC are used for data acquisition.}
    \label{fig:setup}
\end{figure}

Hardware-in-the-loop (HIL) simulations were performed to further validate the proposed control strategy. The plant of \cref{fig:controlDiagram} was modelled in a PLECS RT Box 1, and a Texas Instruments (TI) LaunchPad F28379D microcontroller unit (MCU) was connected externally for control as in \cref{fig:setup}.

\begin{table}[htbp]
    \vspace{-1.4em}
    \centering
    \caption{HIL Simulation Parameters.}
    \vspace{-0.7em}
    \label{tab:RTsimParams}
    \renewcommand{\arraystretch}{1.05}
    \small
    \begin{tabular}{
        >{\raggedright\arraybackslash}p{4.0cm} c @{\hspace{0.5cm}} c
        }
        \toprule
        \textbf{Parameter}           & \textbf{Symbol}     & \textbf{Value}                             \\
        \midrule
        \multicolumn{3}{l}{\textbf{System}}                                                             \\
        \addlinespace
        DC Voltage                   & $V_\text{dc}=V_o$   & \SI{400}{\volt}                            \\
        Output Capacitance           & $C_o$               & \SI{22}{\micro\farad}                      \\
        RT Box Timestep              & $\Delta t$          & \SI{2}{\micro\second}                      \\
        Controller Integral Gain     & $K_i$               & \num{50}\si{\degree\per\ampere\per\second} \\
        \midrule
        \multicolumn{3}{l}{\textbf{Onboard Charger (OBC)}}                                              \\
        \addlinespace
        Switching Frequency          & $f_\text{sw,obc}$   & \SI{25}{\kilo\hertz}                       \\
        Turns Ratio                  & $\alpha_\text{obc}$ & 1                                          \\
        Leakage Inductance           & $L_\text{obc}$      & \SI{136}{\micro\henry}                     \\
        Compensation Capacitance     & $C_\text{obc}$      & \SI{540}{\nano\farad}                      \\
        Controller Proportional Gain & $K_p$               & \num{0.5}\si{\degree\per\ampere}           \\
        \midrule
        \multicolumn{3}{l}{\textbf{Wireless Charger (WC)}}                                              \\
        \addlinespace
        Switching Frequency          & $f_\text{sw,wc}$    & \SI{21.25}{\kilo\hertz}                    \\
        Turns Ratio                  & $\alpha_\text{wc}$  & 2                                          \\
        Leakage Inductance           & $L_\text{wc}$       & \SI{873.6}{\micro\henry}                   \\
        Compensation Capacitance     & $C_\text{wc}$       & \SI{70}{\nano\farad}                       \\
        Controller Proportional Gain & $K_p$               & \num{0.2}\si{\degree\per\ampere}           \\
        \bottomrule
    \end{tabular}
\end{table}

To meet the hardware usage limitations of the RT Box, the switching frequencies were reduced to one quarter of their offline simulation values, while $L_\text{obc}$, $C_\text{obc}$, $L_\text{wc}$, and $C_\text{wc}$ were quadrupled to keep the same impedance. These adjusted values are not suitable for a practical implementation; a lower switching frequency increases component cost, reducing the benefits of integration. This simulation shows that the proposed modulation and control work in a real-time setting, while a practical implementation would use values closer to those in the offline simulation. The HIL simulation parameters are listed in \cref{tab:RTsimParams}, and the HIL simulation results are shown in \cref{fig:Plots_OBC_RT} and \cref{fig:Plots_WC_RT}.

\begin{figure}[htbp]
    \centering
    \subfloat[Transient responses: (i) power tracking and (ii) phase shift.\label{fig:RT_OBC_transient}]{%
        \begin{minipage}{1\linewidth}
            \centering
            \includegraphics[width=1\linewidth]{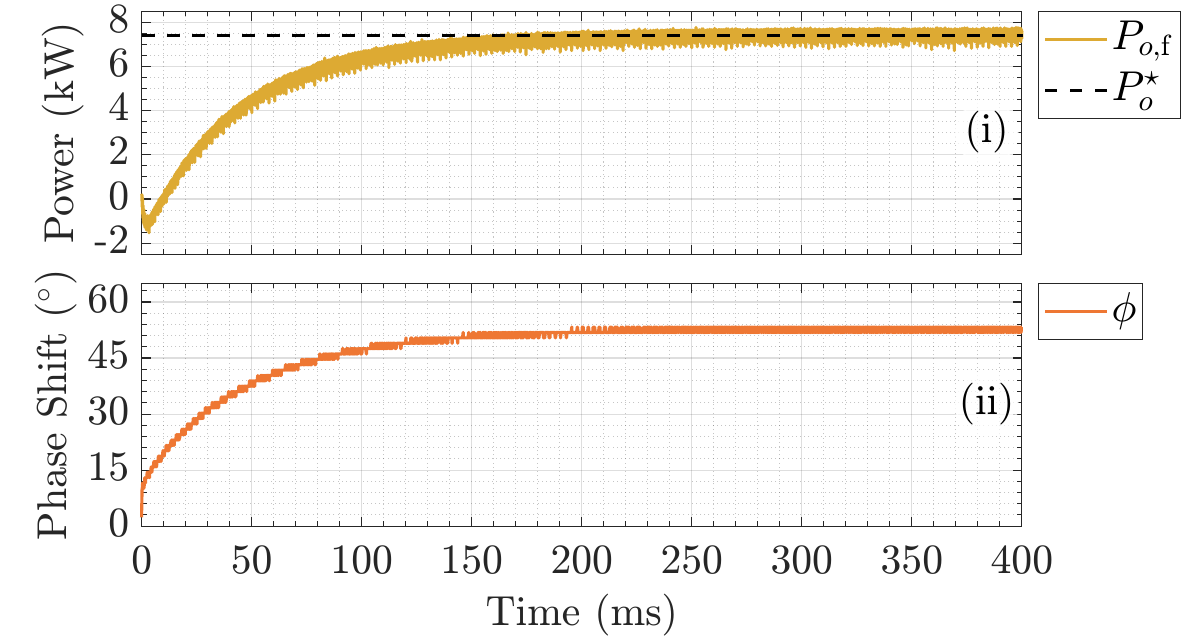}
        \end{minipage}%
    }
    \par
    \subfloat[Decoupled operation: (i) OBC voltages, (ii) WC voltages, (iii) OBC full-bridge gating, and (iv) shared full-bridge gating.\label{fig:RT_OBC_quick}]{%
        \begin{minipage}{1\linewidth}
            \centering
            \includegraphics[width=1\linewidth]{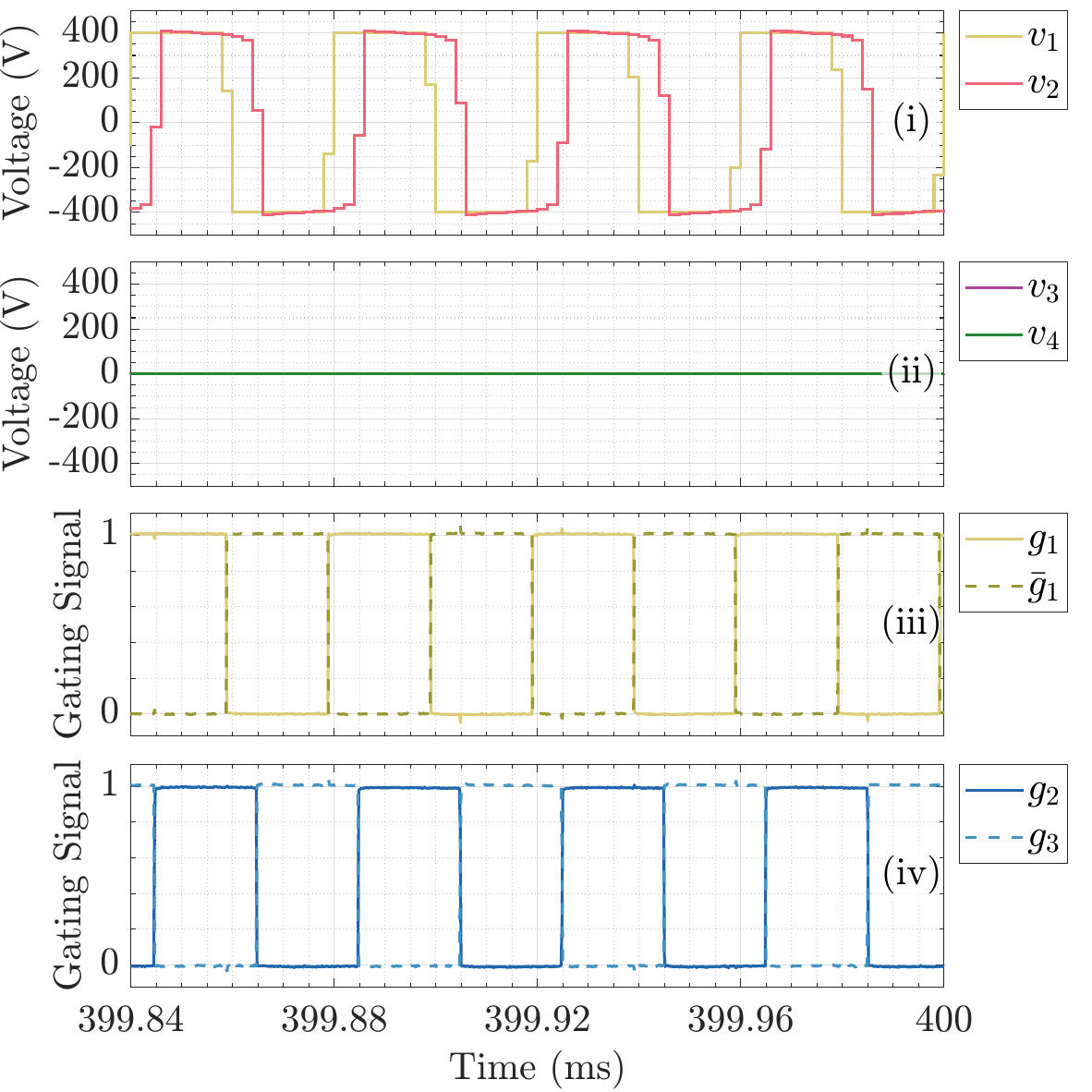}
        \end{minipage}%
    }
    \par
    \IEEEleftcaption{HIL simulation results tracking $P_o^\star=\SI{7.4}{\kilo\watt}$ in OBC mode.}{fig:Plots_OBC_RT}
    \vspace{-1.6em}
\end{figure}

\begin{figure}[htbp]
    \centering
    \subfloat[Transient responses: (i) power tracking and (ii) phase shift.\label{fig:RT_WC_transient}]{%
        \begin{minipage}{1\linewidth}
            \centering
            \includegraphics[width=1\linewidth]{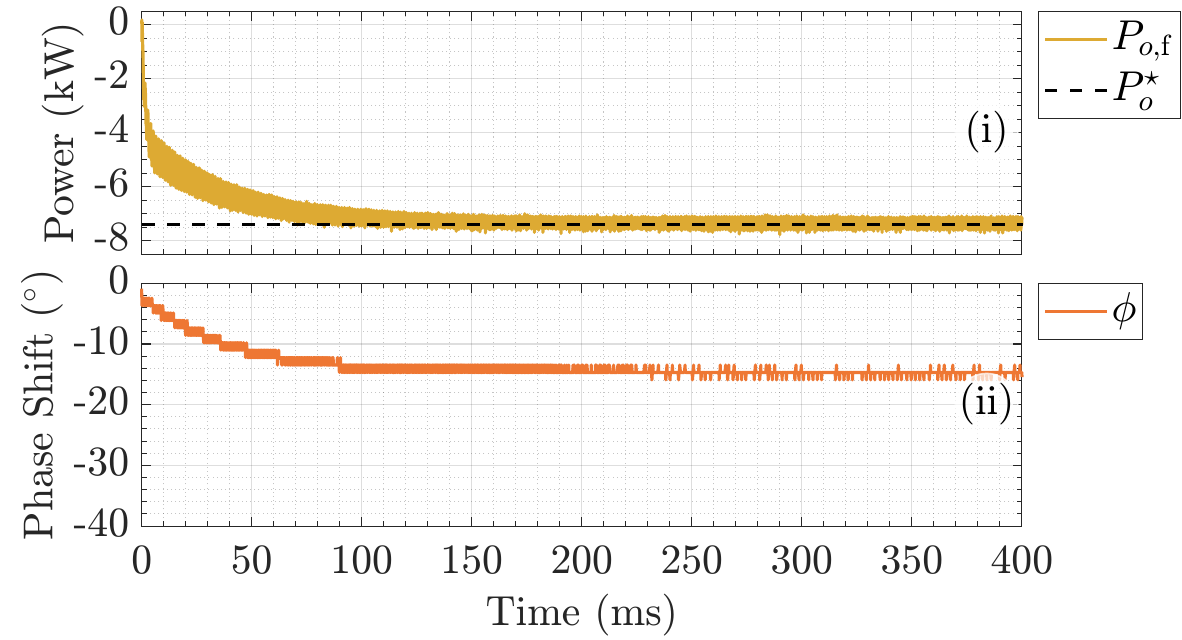}
        \end{minipage}%
    }
    \par
    \subfloat[Decoupled operation: (i) OBC voltages, (ii) WC voltages, (iii) WC full-bridge gating, and (iv) shared full-bridge gating.\label{fig:RT_WC_quick}]{%
        \begin{minipage}{1\linewidth}
            \centering
            \includegraphics[width=1\linewidth]{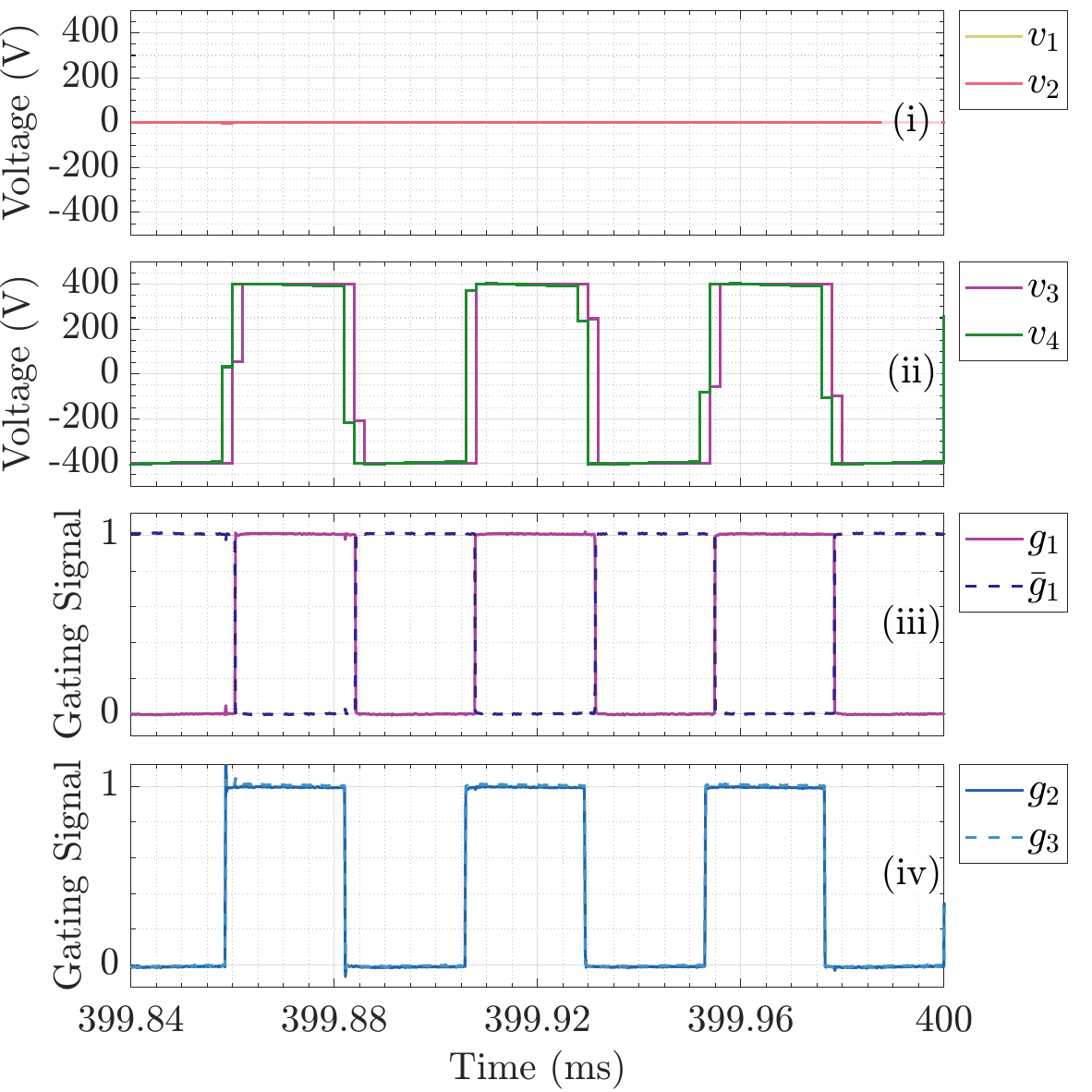}
        \end{minipage}%
    }
    \par
    \IEEEleftcaption{HIL simulation results tracking $P_o^\star=\SI{-7.4}{\kilo\watt}$ in WC mode.}{fig:Plots_WC_RT}
    \vspace{-1.6em}
\end{figure}

The HIL simulation results confirm the expected power tracking behaviour with several implementation-related artifacts. \cref{fig:RT_OBC_transient} (i) and \cref{fig:RT_WC_transient} (i) show that the controller tracks $P_o^\star$ for OBC grid-to-vehicle and WC vehicle-to-grid operation, despite noise in $P_{o,\mathrm{f}}$ introduced by the external connection between the RT Box and the MCU. For positive power references, there is undershoot in $P_{o,\mathrm{f}}$ that is not present in the offline simulation. This initial transient also occurs for negative power references, though it is less obvious because it initially drives the output power toward the negative reference. Since this transient occurs in the negative direction for positive and negative references, it is likely a discretization artifact rather than an inherent feature of the control scheme.

The phase shift in \cref{fig:RT_OBC_transient} (ii) and \cref{fig:RT_WC_transient} (ii) is discretized due to the way it was computed for plotting post hoc. $\phi$ could not be read directly, so it was computed from the time difference between the rising edges of $g_1$ and $g_2$ measured on an oscilloscope. The oscilloscope's sampling interval of \SI{0.16}{\micro\second} gives an angle resolution of \SI{1.44}{\degree} for OBC mode and \SI{1.224}{\degree} for WC mode, hence the obvious discrete steps in \cref{fig:RT_OBC_transient} (ii) and \cref{fig:RT_WC_transient} (ii). The RT Box is not subject to a phase-shift resolution this coarse; rather, the actual resolution is \SI{0.09}{\degree} for OBC mode and \SI{0.0765}{\degree} for WC mode as limited by the MCU. The steady-state phase shifts are also different compared to the offline simulation. For OBC mode, it is \SI{52.5}{\degree}, which is larger than the offline simulation value of \SI{32}{\degree}; for WC mode, it is \SI{-14.5}{\degree}, which is smaller in magnitude than the offline simulation value of \SI{-32}{\degree}. It was found that increasing the time step $\Delta t$ increased the OBC steady-state phase shift but decreased the magnitude of the WC steady-state phase shift. This suggests that decreasing $\Delta t$ would bring the real-time phase shifts closer to their offline simulation values.

The voltages in \cref{fig:RT_OBC_quick} (i) and \cref{fig:RT_WC_quick} (ii) show the expected square-wave behaviour, but the transitions from positive to negative voltage include an extra discrete voltage step. This is a modelling artifact of the RT Box and would not occur in a practical implementation. The full-bridge component in PLECS uses a sub-cycle average of the gating signal to compute its voltage. When the gating signal transitions from high to low, the average is computed over a time window that includes both high and low states, resulting in a voltage that is between the high and low states. The height of the extra voltage step depends on how long the voltage is high or low during the averaging window. Besides this artifact, \cref{fig:RT_OBC_quick} (ii) shows that the WC-side voltages are negligible ($<\SI{1}{\volt}$) in OBC mode, and \cref{fig:RT_WC_quick} (i) shows that the OBC-side voltages are negligible in WC mode, validating the proposed modulation method for decoupling OBC and WC operation.

\section{Conclusion and Future Work}
This paper presents a modulation scheme to unlock the hardware-reduction advantages of the DAB-based integrated OBC/WC proposed in \cite{patent}. This modulation scheme allows bidirectional power transfer for wired and wireless charging, an important feature for vehicle-to-everything applications. The proposed method operates the shared full-bridge as a differential-mode voltage source in OBC mode and as a common-mode voltage source in WC mode. As a result, the active charging path can transfer power while the inactive charger is de-energized, realizing bidirectional wired and wireless charging without additional contactors.

Analytical derivations showed that setting $g_3 = \bar{g}_2$ produces the required OBC voltage while suppressing the WC-side voltage, whereas setting $g_3 = g_2$ produces the required WC voltage while suppressing the OBC-side voltage. Based on these relationships, practical design guidelines are given for selecting the OBC transformer turns ratio and equivalent turns ratio of the WC coils. Offline simulations verified the proposed method by demonstrating charger decoupling in both charging modes. HIL simulations using a PLECS RT Box and an external TI MCU further showed that the modulation scheme can be implemented in a real-time control environment. Although discretization introduced visible differences compared to the offline simulation, the results confirmed the intended voltage decoupling and power transfer behaviour.

Future work will focus on a hardware implementation that uses the proposed modulation strategy. This will include bidirectional wired and wireless power-transfer tests on a physical prototype, evaluation of the decoupling performance, and an efficiency comparison with other non-integrated and integrated EV charger designs.

\bibliographystyle{IEEEtran}
\bibliography{IEEEabrv,bibliography}

\end{document}